\documentclass[%
 reprint,
 amsmath,amssymb,
 aps,
]{revtex4-2}

\usepackage{graphicx} 
\usepackage{dcolumn}
\usepackage{bm} 

\usepackage[colorlinks=true,linkcolor=blue,citecolor=blue,urlcolor=blue]{hyperref}
\usepackage{url}

\usepackage{amsfonts}
\usepackage{amssymb}
\usepackage{graphics}
\usepackage{epstopdf}
\usepackage{longtable}
\usepackage{epsfig}
\usepackage{times}
\usepackage{color}
\usepackage{xcolor}
\usepackage[table]{xcolor}
\usepackage{float}
\usepackage{siunitx}
\usepackage{orcidlink}
\usepackage{tikz}
\usepackage{CJK}
\usepackage{amsmath}
\usetikzlibrary{shapes.geometric, arrows, positioning}

\begin{document}

\title{Enhanced Yield Rate of \textsuperscript{229m}Th via Cascade Decay in Storage Rings and Electron Beam Ion Traps}

\author{Yumiao \surname{Wang}\,\orcidlink{0009-0006-8140-2424}}
\author{Yi \surname{Yang}\,\orcidlink{0000-0001-7275-3982}}
\author{Yixin \surname{Li}\,\orcidlink{0000-0002-9815-4272}}
\author{Ding \surname{Yue}\,\orcidlink{0009-0002-6801-2668}}
\author{Kai \surname{Zhao}\,\orcidlink{0009-0000-0821-9697}}
\author{Youjing \surname{Wang}\,\orcidlink{0000-0002-5260-1823}}
\author{Changbo \surname{Fu}\,\orcidlink{0000-0001-5895-7823}}
\email{cbfu@fudan.edu.cn}
\author{Yugang \surname{Ma}\,\orcidlink{0000-0002-0233-9900}}
\hypersetup{urlcolor=black}

\affiliation{Key Laboratory of Nuclear Physics and Ion-beam Application (MoE), Institute of Modern Physics, Fudan University, Shanghai 200433, China}

\date{\today}

\begin{abstract}
The low-energy nuclear isomeric state of \textsuperscript{229m}Th provides a unique bridge between nuclear and atomic physics, enabling applications such as nuclear clocks and precision metrology.
However, efficient and controllable production of \textsuperscript{229m}Th remains a major experimental challenge. 
We propose an efficient scheme to produce the $^{229\mathrm{m}}$Th in storage rings (SRs) and electron beam ion traps (EBITs), using a cascade decay pathway. Highly charged ions are excited to higher nuclear states via nuclear excitation by inelastic electron scattering (NEIES) and nuclear excitation by electron capture (NEEC), followed by radiative or internal conversion cascades that populate the isomer.
Our calculations demonstrate that, under typical SRs and EBITs conditions, optimized indirect excitation pathways significantly enhance \textsuperscript{229m}Th production rate. 
In particular, NEIES can provide an enhancement of up to four orders of magnitude through cascade de-excitation at high energies, while NEEC can contribute an additional enhancement of up to  several tens of times.
Such a significant increase in the \textsuperscript{229m}Th yield rate would facilitate its application in various nuclear photonics fields, especially in the development of atomic nuclear clocks.

\end{abstract}

\maketitle

\section{Introduction}

The low-lying isomeric state of \textsuperscript{229m}Th at  $8.36$~eV~\cite{KROGER197629, PhysRevLett.64.271} has attracted intense interest due to its potential applications in nuclear optical clocks~\cite{Tkalya1996, Peik2003, vonDerWense2020, Thirolf2024, Kraemer2025}, nuclear lasers~\cite{PhysRevLett.106.162501}, tests of fundamental constants~\cite{Peik2021, Beeks2025}, searching for dark matter~\cite{PhysRevX.15.021055}, and nuclear-based frequency standards~\cite{zhang2024frequency}. 

Despite this remarkable progress in spectroscopic characterization~\cite{tiedau2024laser, elwell2024laser, zhang2024frequency, Kraemer2023, Hiraki2024, Pineda2025, Morgan2025, Masuda2025}, 
the efficient and controllable production of \textsuperscript{229m}Th remains one of the key challenges for both crystal-based~\cite{tiedau2024laser, elwell2024laser, zhang2024frequency, 
zhang2024229thf4,
hiraki2024controlling, Elwell2025} 
and ion-trap-based~\cite{yamaguchi2024laser, Zitzer2024, zitzer2025laser, ztxz-dwhk, moritz2025cryogenic} applications.
\textsuperscript{229m}Th can be produced through the $\alpha$ decay of \textsuperscript{233}U~\cite{von2016direct, seiferle2017lifetime,
thielking2018laser, seiferle2019energy, yamaguchi2019energy, yamaguchi2024laser} or the $\beta$ decay of \textsuperscript{229}Ac~\cite{kraemer2023observation, PhysRevC.100.024315}, but with very limited production rates.
Direct laser excitation in Th-doped crystals has been demonstrated~\cite{tiedau2024laser, elwell2024laser, zhang2024frequency}, but the excitation efficiency remains extremely low, and high-power continuous-wave VUV sources are still under development~\cite{thielking2023vacuum, xiao2024proposal, lal2025continuous}. 
Alternative indirect schemes via higher-lying nuclear states, including the second excited state at 29.19~keV, have been explored using synchrotron radiation~\cite{masuda2019x}, yet they require large-scale facilities and offer limited tunability.

To produce \textsuperscript{229m}Th with an efficient mechanism, 
electron-mediated nuclear excitation may provide a versatile and controllable alternative, with two mechanisms being particularly relevant: nuclear excitation by inelastic electron scattering (NEIES, also known as Coulomb excitation, CE)~\cite{temmer1956contribution, tkalya2020excitation, zhang2023theory, PhysRevC.106.044604, PhysRevC.110.064621} and nuclear excitation by electron capture (NEEC, or inverse internal conversion, IIC)~\cite{palffy2006theory, palffy2007isomer, chiara2018isomer, gargiulo2022nuclear, wang2023feasibility, zhang2023theory, zhao2024efficient, yang2024feasibility, yang2025physics, xu2025charge}. 
In NEIES, free electrons transfer part of their kinetic energy to the nucleus through inelastic collisions whenever the electron energy exceeds the excitation threshold. 
NEEC, in contrast, is a resonant process in which a free electron is captured into a bound electronic state of an ion while simultaneously exciting the nucleus, enabling selective and efficient population of nuclear states. 
The \textsuperscript{229}Th isomer has been studied theoretically under the NEIES and NEEC mechanisms~\cite{zhao2024efficient, PhysRevC.110.064621, xu2025charge}, 
focusing on its first and second excited states.
Experimentally, electron-mediated excitations of \textsuperscript{229}Th have not yet been realized~\cite{tkalya2020excitation, PhysRevLett.127.052501, PhysRevC.110.064621, PhysRevC.111.064302}. In addition, the NEEC process has not yet been confirmed~\cite{chiara2018isomer, wu2019mo, PhysRevLett.128.242502, PhysRevLett.127.042501, PhysRevC.108.L031302}. 
Both processes are significantly enhanced in highly charged ions (HCIs)~\cite{palffy2006theory, kozlov2018highly, tu2025probing}, where reduced electronic shielding strengthens the electron-nucleus coupling, making HCIs ideal for exploiting controlled excitation pathways. 

\begin{figure*}[!htbp]
\centering
\begin{tikzpicture}[node distance=1cm, >=stealth]

\node at (-8.3, -0.8) {$\beta$};
\node at (-7.8, -0.4) {$J^{\pi}$};
\node at (8.5, -0.4) {$T_{1/2}$};
\node at (8.5, -0.8) {$E_n$~(keV)};

\draw[line width=0.8mm, color=black] (-8, 0) -- (7.8, 0); 
\draw[line width=0.8mm, color=red] (-8, 1) -- (7.8, 1); 
\draw[line width=0.8mm, color=cyan] (-8, 2) -- (7.8, 2);  
\draw[line width=0.8mm, color=cyan] (-8, 3) -- (7.8, 3);   
\draw[line width=0.8mm, color=cyan] (-8, 4) -- (7.8, 4);  
\draw[line width=0.8mm, color=cyan] (-8, 5) -- (7.8, 5); 
\draw[line width=0.8mm, color=cyan] (-8, 6) -- (7.8, 6); 

\node at (-8.3, 0) {0};
\node at (-8.3, 1) {1};
\node at (-8.3, 2) {2};
\node at (-8.3, 3) {3};
\node at (-8.3, 4) {4};
\node at (-8.3, 5) {5};
\node at (-8.3, 6) {6};

\node at (-7.8, 0.3) {5/2$^+$};
\node at (-7.8, 1.3) {3/2$^+$};
\node at (-7.8, 2.3) {5/2$^+$};
\node at (-7.8, 3.3) {7/2$^+$};
\node at (-7.8, 4.3) {7/2$^+$};
\node at (-7.8, 5.3) {9/2$^+$};
\node at (-7.8, 6.3) {9/2$^+$};

\node at (8.5, 0.3) {7880 y};
\node at (8.5, 1.3) {7 $\mu$s};
\node at (8.5, 2.3) {82.2 ps};
\node at (8.5, 3.3) {0.172 ns};
\node at (8.5, 4.3) { };
\node at (8.5, 5.3) {0.147 ns};
\node at (8.5, 6.3) { };

\node at (8.5, 0.0) {0.0};
\node at (8.5, 1.0) {0.00836};
\node at (8.5, 2.0) {29.19};
\node at (8.5, 3.0) {42.43};
\node at (8.5, 4.0) {71.83};
\node at (8.5, 5.0) {97.14};
\node at (8.5, 6.0) {125.44};

\draw[->, blue, thick] (-7.3, 0) -- (-7.3, 6) node[below right, yshift=-3, xshift=-0.5] {E2};
\node at (-7.3, 6) [circle, draw=blue, inner sep=0.7mm, line width=0.3mm] {};
\draw[->, blue, thick] (-6.3, 0) -- (-6.3, 5) node[below right, yshift=-3, xshift=-0.5] {E2};
\node at (-6.3, 5) [circle, draw=blue, inner sep=0.7mm, line width=0.3mm] {};
\draw[->, brown, thick] (-5.3, 0) -- (-5.3, 4) node[below right, yshift=-3, xshift=-0.5] {M1+E2};
\node at (-5.3, 4) [circle, draw=brown, inner sep=0.7mm, line width=0.3mm] {};
\draw[->, brown, thick] (-4.3, 0) -- (-4.3, 3) node[below right, yshift=-3, xshift=-0.5] {M1+E2};
\node at (-4.3, 3) [circle, draw=brown, inner sep=0.7mm, line width=0.3mm] {};
\draw[->, brown, thick] (-3.3, 0) -- (-3.3, 2) node[below right, yshift=-3, xshift=-0.5] {M1+E2};
\node at (-3.3, 2) [circle, draw=brown, inner sep=0.7mm, line width=0.3mm] {};
\draw[->, brown, thick] (-2.3, 0) -- (-2.3, 1) node[below right, yshift=-3, xshift=-0.5] {M1+E2};
\node at (-2.3, 1) [circle, draw=brown, inner sep=0.7mm, line width=0.3mm] {};

\node at (-1.0, 6) [circle, draw=blue, inner sep=0.7mm, line width=0.3mm, scale=1] {};
\draw[->, blue, thick] (-1.0, 6) -- (-1.0, 2) node[above right, yshift=3, font=\scriptsize, xshift=-0.5] {21.0};

\node at (0.0, 6) [circle, draw=brown, inner sep=0.7mm, line width=0.3mm, scale=1] {};
\draw[->, brown, thick] (0.0, 6) -- (0.0, 4) node[above right, yshift=3, font=\scriptsize, xshift=-0.5] {72.0};

\node at (1.0, 6) [circle, draw=brown, inner sep=0.7mm, line width=0.3mm, scale=1] {};
\draw[->, brown, thick] (1.0, 6) -- (1.0, 5) node[above right, yshift=3, font=\scriptsize, xshift=-0.5] {4.6};

\node at (2.0, 5) [circle, draw=blue, inner sep=0.7mm, line width=0.3mm, scale=1] {};
\draw[->, blue, thick] (2.0, 5) -- (2.0, 2) node[above right, yshift=3, font=\scriptsize, xshift=-0.5] {1.4};

\node at (3.0, 5) [circle, draw=brown, inner sep=0.7mm, line width=0.3mm, scale=1] {};
\draw[->, brown, thick] (3.0, 5) -- (3.0, 4) node[above right, yshift=3, font=\scriptsize, xshift=-0.5] {27.1};

\node at (4.0, 4) [circle, draw=blue, inner sep=0.7mm, line width=0.3mm, scale=1] {};
\draw[->, blue, thick] (4.0, 4) -- (4.0, 1) node[above right, yshift=3, font=\scriptsize, xshift=-0.5] {11.8};

\node at (5.0, 4) [circle, draw=brown, inner sep=0.7mm, line width=0.3mm, scale=1] {};
\draw[->, brown, thick] (5.0, 4) -- (5.0, 2) node[above right, yshift=3, font=\scriptsize, xshift=-0.5] {73.5};


\node at (6.0, 2) [circle, draw=brown, inner sep=0.7mm, line width=0.3mm, scale=1] {};
\draw[->, brown, thick] (6.0, 2) -- (6.0, 1) node[above right, yshift=3, font=\scriptsize, xshift=-0.5] {58}; 

\node at (7.0, 1) [circle, draw=brown, inner sep=0.7mm, line width=0.3mm, scale=1] {};
\draw[->, brown, thick] (7.0, 1) -- (7.0, 0) node[above right, yshift=3, font=\scriptsize, xshift=-0.5] {100};

\end{tikzpicture}
\caption{\label{fig1-levels}
Partial level scheme of \textsuperscript{229}Th showing selected nuclear excited states and their decay pathways to the isomeric state. 
The M1+E2 and E2 transitions in the decay paths are marked in brown and blue, respectively. 
For each nuclear level
$\beta$, the spin--parity assignment $J^{\pi}$, half-life $T_{1/2}$, and excitation energy $E_n$ are labeled on either side of the corresponding level line. 
The upper-right label on each arrow indicates the branching ratio of the corresponding decay channel (in \%). 
The energy and lifetime of the first excited state are taken from~\cite{PhysRevLett.106.162501}, and those of the second excited state are also taken from~\cite{masuda2019x}. 
For the branching ratios of different decay channels, the values for the second excited state are taken from~\cite{masuda2019x}, while the others are calculated using the formulas Eqs.~(\ref{eq_BR}), with the relevant data obtained from~\cite{NNDC}. 
Unless otherwise specified, all other data are also taken from~\cite{NNDC}.
}
\end{figure*}
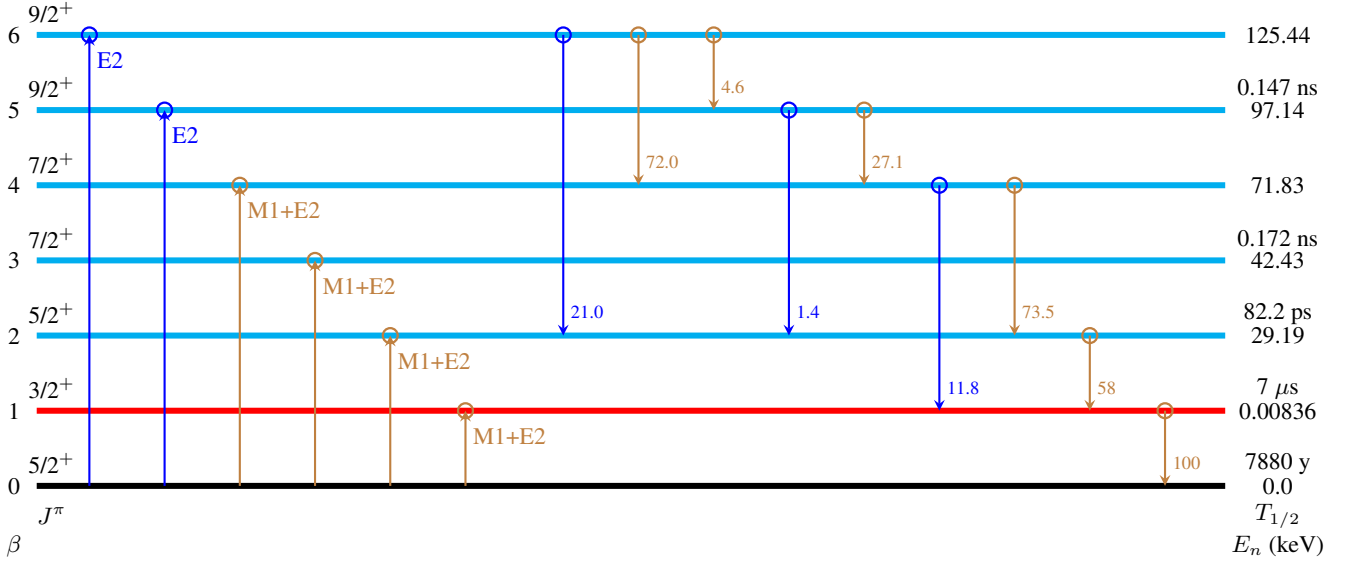

There are two experimental platforms for producing HCIs, storage rings (SRs)~\cite{bosch2013nuclear, ma2015proposal, yang2024feasibility, jin2023excitation} and electron beam ion traps (EBITs)~\cite{fu2010overview, wang2023feasibility, tu2025probing, prokhorov2025coulomb}, which are suitable for \textsuperscript{229m}Th producing applications \cite{zhao2024efficient}. 
In SRs, circulating HCIs interact with an electron cooler or a dedicated electron target, enabling nuclear excitation and downstream detection of isomers. 
In EBITs, HCIs are produced and confined by a compressed, tunable electron beam combined with magnetic and electrostatic fields.
In these apparatuses, the NEIES and NEEC processes dominate the production of \textsuperscript{229m}Th \cite{qi2023isomeric, zhao2024efficient}, making them critical to the design of electron-driven isomeric excitation schemes. 
NEIES occurs whenever the electron energy exceeds the excitation threshold, while NEEC can be resonantly induced.

The selected excitation scheme to higher levels, followed by cascade decay to the \textsuperscript{229}Th isomeric state, is shown in Fig.~\ref{fig1-levels}.
As discussed by Chen et al.~\cite{chen2025microscopic}, by using the projected shell model, some nuclear transition probabilities of higher-lying states of \textsuperscript{229}Th are relatively large.
Therefore, the cascade decay to the isomeric state can be utilized to optimize isomer yield, and then may achieve significantly higher efficiency compared to direct excitation methods.

In this work, we present a theoretical investigation of electron-mediated excitation pathways leading to \textsuperscript{229m}Th via cascade decay under SR and EBIT conditions. 
We calculate NEIES and NEEC cross sections, excitation rates, and their dependence on electron energy, charge state, and electronic configuration, providing quantitative guidance for optimizing isomer production and informing the design of future nuclear-clock and NEEC experiments. 

The remainder of this paper is organized as follows. 
Section~\ref{sec_theory} introduces the theoretical framework for NEIES and NEEC, as well as the procedures used to compute electronic wave functions.
Section~\ref{sec_results} presents energy-dependent cross sections and production rates for different pathways under EBIT and SR conditions. 
The main results and an outlook for experimental implementations will be presented at the end.

\section{Theoretical framework of electronic processes}\label{sec_theory}

In the SR and EBIT environments, the isomeric producing mechanisms are the NEIES and NEEC, which are briefly introduced in the following sections.
Throughout this paper, atomic units (a.u.) are employed, such that $\hbar = m_{e} = e = 1$ and $c = 1/\alpha$. In these units, $\hbar$ represents the reduced Planck constant, $m_{e}$ the mass of the electron, $e$ the elementary charge, $c$ the speed of light, and $\alpha$ the fine-structure constant.

\subsection{NEIES cross section}

The detailed calculations of NEIES and NEEC have been extensively discussed in previous works~\cite{zhang2023theory}. Therefore, we present only the final expression for the NEIES cross section, which is expressed as 
\begin{equation}
\label{eq_sigma_neies}
\begin{aligned}
\sigma_{\mathrm{NEIES}}(E_i) =& \frac{8\pi^2}{c^4} \frac{E_i + c^2}{p_i^3} \frac{E_f + c^2}{p_f} \\
& \times \sum_{T,\lambda} \frac{k^{2\lambda+2}}{(2\lambda-1)!!^2} B\left(T\lambda, J_i \rightarrow J_f\right) \\
& \times \sum_{l_i, j_i, l_f, j_f} \frac{(2l_i+1)(2l_f+1)(2j_i+1)(2j_f+1)}{(2\lambda+1)^2} \\
& \times \left( \begin{array}{ccc} l_f & l_i & \lambda \\ 0 & 0 & 0 \end{array} \right)^2 \left\{ \begin{array}{ccc} l_i & \lambda & l_f \\ j_f & 1/2 & j_i \end{array} \right\}^2 \left| R_{f i}^{T\lambda} \right|^2.
\end{aligned}
\end{equation}
Here, the subscripts $i$ and $f$ label the initial and final electronic states, while the indices $n$ and $e$ distinguish nuclear and electronic quantities. In Eq.~(\ref{eq_sigma_neies}), $E_i$ and $E_f$ denote the energies of the incident and outgoing electrons, with corresponding momenta $p_i$ and $p_f$. The factor $k = E_{n}/c$ represents the nuclear excitation energy divided by the speed of light, while $T$ and $\lambda$ specify the type (electric or magnetic) and multipolarity of the nuclear transition. $B\left(T\lambda, J_i \rightarrow J_f\right)$ is the reduced nuclear transition probability characterizing the intrinsic strength of the nuclear transition between initial and final nuclear states with spins $J_i$ and $J_f$. The sums over $l_i$, $j_i$, $l_f$, and $j_f$ account for the possible angular momentum couplings of the electronic states, with $n$, $l$, $j$, and $m$ denoting the principal quantum number, orbital angular momentum, total angular momentum, and magnetic quantum number of the initial and final electronic states, respectively. 

The radial transition matrix element $R_{fi}^{T\lambda}$ measures the overlap between the initial and final electronic Dirac wavefunctions, representing the electronic contribution to the interaction between the electron and the nucleus. The same formalism is, in principle, also applicable to the NEEC process. For electric transitions, one has  
\begin{equation}
\label{eq_te}
R_{f i}^{E\lambda} = \int_0^\infty h_\lambda^{(1)}(k r) \left[ g_i(r)g_f(r) + f_i(r)f_f(r) \right] r^2 dr,
\end{equation}  
and for magnetic transitions,   
\begin{equation}
\label{eq_tm}
\begin{aligned}
R_{f i}^{M\lambda} =& \frac{\kappa_i + \kappa_f}{\lambda} \\
& \times \int_0^\infty h_\lambda^{(1)}(k r) \left[ g_i(r)f_f(r) + g_f(r)f_i(r) \right] r^2 dr.
\end{aligned}
\end{equation}  
Here, $\kappa = (l - j)(2j + 1)$ is conventionally used as short-hand notation for $j$ and $l$, and $h_\lambda^{(1)}(k r)$ represents the spherical Hankel function of the first kind. 

The radial components of the electronic wave functions, $g(r)$ and $f(r)$, correspond to the initial and final states, respectively. In this work, both bound and continuum Dirac wave functions are computed using the \texttt{RADIAL} package~\cite{salvat2019radial}. The underlying electron potential is obtained within the Dirac-Hartree-Fock-Slater framework~\cite{PhysRev.81.385, LIBERMAN1971107}, which incorporates the nuclear, direct electronic, and exchange contributions in a self-consistent manner. The detailed formulation of the DHFS potential can be found in Refs.~\cite{PhysRevC.110.064621, xu2025charge}. 

\subsection{NEEC cross section}

The NEEC cross section~\cite{zhang2023theory} can be given by 
\begin{equation}
\label{eq_sigma_neec}
\begin{aligned}
\sigma_{\mathrm{NEEC}}(E_{i})= & \frac{8 \pi^3}{c^2} \frac{E_i+c^2}{p_i^3} \\
 & \times \sum_{T,\lambda}  \frac{k^{2 \lambda+2}}{(2 \lambda-1)!!^2} B\left(T\lambda, J_i \rightarrow J_f\right) \\
& \times  \sum_{l_i, j_i, l_f, j_f} \frac{\left(2 l_i+1\right)\left(2 l_f+1\right)\left(2 j_i+1\right)\left(2 j_f+1\right)}{(2 \lambda+1)^2} \\
& \times\left(\begin{array}{ccc}
l_f & l_i & \lambda \\
0 & 0 & 0
\end{array}\right)^2\left\{\begin{array}{ccc}
l_i & \lambda & l_f \\
j_f & 1 / 2 & j_i
\end{array}\right\}^2\left|R_{f i}^{T \lambda}\right|^2 \\
& \times L_{\mathrm{NEEC}} (E_{i}-E_{r}).
\end{aligned}
\end{equation} 
Here, $E_r$ represents the resonance energy of the recombining electron. The function $L_{\mathrm{NEEC}}$ represents the normalized Lorentz profile occurring in resonant systems,
\begin{equation}
L_{\mathrm{NEEC}} (E_i-E_{r} )=\frac{\Gamma_{\mathrm{NEEC}}/2\pi }{(E_i-E_{r} )^{2}+\Gamma_{\mathrm{NEEC}}^{2}/4}.
\end{equation}
In the above expression, the natural width of the resonant state, $\Gamma_{\mathrm{NEEC}}$, consists of two main contributions: the width of the final electronic state, $\Gamma_{f}$, associated with spontaneous photon emission, and the nuclear decay width, $\Gamma_{n} = \Gamma_{\gamma} + \Gamma_{\mathrm{IC}}$, which includes both the $\gamma$-decay width, $\Gamma_{\gamma}$, and the internal conversion width, $\Gamma_{\mathrm{IC}}$. When the electron is captured directly into the ionic ground state, the total width reduces to $\Gamma_{\mathrm{NEEC}} = \Gamma_{n}$. In contrast, if the capture occurs into an excited electronic state, the total width becomes $\Gamma_{\mathrm{NEEC}} = \Gamma_{n} + \Gamma_{f}$.

In cases where the incident electron energy is distributed rather than sharply defined, an integration over the electron energy becomes necessary. To simplify the treatment of this effect, the concept of resonant strength is employed as \begin{equation}
\label{eq_S_neec}
S_{\mathrm{NEEC}}=\int d E_i \sigma_{\mathrm{NEEC}}\left(E_i\right).
\end{equation}

\subsection{Excitation rate}
For a given excitation channel $0\rightarrow\mathrm{\beta}$, the NEIES and NEEC excitation rates per target particle are obtained by convoluting the corresponding electric and magnetic transition cross sections with the electron flux energy distribution $\Phi_e(E_i)$ over the incident electron energy $E_i$. They are calculated by
\begin{equation}
\label{eq_per_rate}
\begin{split}
\lambda_{{\mathrm{NEIES/NEEC}}} = \int dE_i  \sigma_{\mathrm{NEIES/NEEC}}(E_i)  \Phi_e(E_i).
\end{split}
\end{equation} 
Owing to the extremely narrow width of the NEEC resonance relative to the energy spread of the incident electron, the process can be treated within the resonance approximation, where the energy dependence of the cross section is represented by a Dirac delta function, $\delta(E_i - E_r)$. Specifically, in the case of NEEC, the excitation rate can be equivalently expressed in terms of the resonance strength $S$ as $\lambda_{{\mathrm{NEEC}}}=S_{\mathrm{NEEC}} \Phi_e(E_r)$.

Based on the electron beam parameters, the electron flux $\Phi_e(E_i)$ is modeled as a Gaussian distribution
\begin{equation}
\Phi_e(E_i)=\frac{I_e}{e A_{e} \epsilon_e \sqrt{\pi}} \exp \left[-\left(\frac{E_i-E_{ir}}{\epsilon_e}\right)^2\right].
\end{equation}
Here, $E_{ir}$ is a function of central
electron energy, $I_e = en_ev_e$ is the electron beam current, 
$e$ is the elementary charge ($1.602\times10^{-19}$~C), $\epsilon_e$ is the electron beam energy spread, and $A_e$ is the beam cross-sectional area, determined from a radius of $r_e$. 

For the non-resonant NEIES process, the cross section exhibits only a weak energy dependence within the narrow energy window $\epsilon_e$ defined by the electron energy distribution, and can thus be reasonably approximated as a constant, $\sigma_\mathrm{NEIES}(E_i) \approx \sigma_0$, over the relevant energy range. Under this approximation, the convolution integral reduces to the standard Gaussian normalization. Thus, $\lambda_{\mathrm{NEIES}}$ simplifies to $\lambda_{{\mathrm{NEIES}}}
= I_e/(e\, A_e)\, \sigma_0$.

In general, the characteristic timescales of the nuclear excitation processes are much shorter than the decay lifetime of the isomeric state: the de-excitation of higher-lying excited states to the isomer occurs on the order of nanoseconds, whereas the isomer itself decays to the ground state on the microsecond timescale. Therefore, unlike Ref.~\cite{PhysRevC.110.064621}, we do not employ a dynamical model that simultaneously describes excitation and decay, and instead neglect the temporal evolution of the isomer population in \textsuperscript{229}Th. 

In order to calculate the population rate of the first excited nuclear state, it is necessary to determine the branching ratios (BRs) of higher-lying states decaying into it. The BR of the second excited state is directly taken from the experimental results reported in Ref.~\cite{masuda2019x}, while those of other states are obtained from experimental data compiled in the NNDC database~\cite{NNDC}. The corresponding formulas used for the calculations are presented below. 

Consider an excited state that decays via multiple $\gamma$ transitions or IC, with no other particle decay channels. 
Let $I_{\beta \to \beta'}$ represent the measured $\gamma$-ray intensity of the $\beta \to \beta'$ transition, and let $\alpha_{\beta \to \beta'}^{\mathrm{eff}}$ denote the associated effective internal conversion coefficient (ICC), where $0 \leq \beta' < \beta$.
For a transition with mixed multipolarity (e.g., M1+E2), the effective ICC is
\begin{equation}
\alpha_{\beta\rightarrow\beta'}^{\mathrm{eff}} =
\dfrac{\alpha_{M,{\beta\rightarrow\beta'}} + \delta_{\beta\rightarrow\beta'}^2 \,\alpha_{E,{\beta\rightarrow\beta'}}}{1 + \delta_{\beta\rightarrow\beta'}^2},
\end{equation}
where $\delta_{\beta\rightarrow\beta'}$ denotes the mixing ratio, reflecting the relative strength of the electric and magnetic components, and $\alpha_{T,{\beta\rightarrow\beta'}}$ is the corresponding ICC.
The BR of each decay channel to the isomeric state is defined as the fraction of the total decay strength
\begin{equation}\label{eq_BR}
\mathrm{BR}_{\beta\rightarrow\beta'} = 
\frac{I_{\beta\rightarrow\beta'} \left( 1 + \alpha_{\beta\rightarrow\beta'}^{\mathrm{eff}} \right)}
{\sum_{\beta\rightarrow\beta'} I_{\beta\rightarrow\beta'} \left( 1 + \alpha_{\beta\rightarrow\beta'}^{\mathrm{eff}} \right)}.
\end{equation}
The effective excitation rate $\lambda^{\mathrm{eff}}_{\mathrm{NEIES/NEEC}}$ is determined by taking into account all possible cascade decay paths from $0 \rightarrow \beta \Rightarrow 1$. Here, the symbol $a \Rightarrow b$ denotes a multi-step cascade process from nuclear state $a$ to state $b$, for instance, as illustrated in Fig.~\ref{fig1-levels}. 
For a given incident electron energy, the effective excitation rate can be expressed as a sum over all possible cascade paths from state $0$ to state $1$ via intermediate states $\beta$, weighted by the product of the corresponding branching ratios
\begin{equation}
\lambda^{\mathrm{eff}}_{\mathrm{NEIES/NEEC}} 
= \sum_{0\rightarrow \beta} \prod_{\beta \Rightarrow 1} \mathrm{BR}_{\beta \Rightarrow 1} \, \lambda_{\mathrm{NEIES/NEEC}}.
\end{equation}
Here, the range of $\beta$ values included depends on whether the corresponding electron energies exceed the energy of the respective nuclear excited state. 

In the following section, the \textsuperscript{229}Th isomer yield rate $R_{\mathrm{iso}}$ is described by different expressions under different experimental setups. We first define an enhancement factor
$P = R_{\mathrm{iso}}^{0 \rightarrow \beta \Rightarrow 1}/R_{\mathrm{iso}}^{0 \rightarrow 1}$,
which represents the relative yield rate of indirect excitation via higher-lying excited states $0 \rightarrow \beta \Rightarrow 1$ compared to direct excitation $0 \rightarrow 1$.

\section{Optimization Results of $^{229m}$Th Production}
\label{sec_results}
 In the following, the calculated cross sections and corresponding production rates for the SR and EBIT are presented separately, followed by a detailed analysis and discussion.

\subsection{Cross sections calculation}
\begin{table*}[!htbp]
\centering
\caption{
Different nuclear transition channels of \textsuperscript{229}Th, together with the reduced transition probabilities for decays to the ground state (in W.u.). The data in the first five rows are taken from Ref.~\cite{chen2025microscopic}.
}
\label{B_T_lambda}
\begin{ruledtabular}
\begin{tabular}{cccccc}
    $\beta\rightarrow0$ & Level (keV)  & $J_{i}^{\pi}$ & $J_{f}^{\pi}$   & $T\lambda$  & $B\left(T\lambda, J_i \rightarrow J_f\right)$ (W.u.) \\
\hline \\ [-0.5pc]
       $1\rightarrow0$ & 0.00836       &      5/2$^+$              &       3/2$^+$                  & $M1 / E2$       &  0.0240 / 8.74      \\[0.2pc]
       $2\rightarrow0$ & 29.19              &      5/2$^+$              &       5/2$^+$             & $M1 / E2$       &  0.00339 / 8.4      \\[0.2pc]
       $3\rightarrow0$ & 42.43               &      5/2$^+$              &       7/2$^+$            & $M1 / E2$       &  0.005 / 233.0      \\[0.2pc]
       $4\rightarrow0$ & 71.83               &      5/2$^+$              &       7/2$^+$            & $M1 / E2$       &  0.001 / 0.018      \\[0.2pc]
       $5\rightarrow0$ & 97.14              &      5/2$^+$              &       9/2$^+$             & $E2$            &  66.4               \\[0.2pc]
       $6\rightarrow0$ & 125.44              &      5/2$^+$              &       9/2$^+$            & $E2$            &  41.936             \\[0.2pc]
\end{tabular}
\end{ruledtabular}
\end{table*}

To evaluate the NEIES cross section in Eq.~(\ref{eq_sigma_neies}) and the NEEC resonance strength in Eq.~(\ref{eq_S_neec}), reduced transition probabilities $B(T\lambda, J_i \to J_f)$ are required. 
The values adopted in the present calculations are summarized in Table~\ref{B_T_lambda}, where the $B(T\lambda, J_i \to J_f)$ for levels~1--5 are taken from the most recent theoretical calculations reported in Ref.~\cite{chen2025microscopic}. 
For the sixth excited state, the $B\left(T\lambda, J_i \rightarrow J_f\right)$ value is evaluated using the \texttt{TROPIC} program~\cite{lee2025tropic}, based on the experimental lifetime, $\gamma$-ray intensity, and ICC from the NNDC database~\cite{NNDC}. 
Due to the lack of experimental $B\left(T\lambda, I_i \rightarrow I_f\right)$ values, theoretical predictions may vary, causing uncertainties in the cross sections~\cite{xiao2025cascading}.

\begin{figure}[!htbp]
\centering
\includegraphics[width=\linewidth]{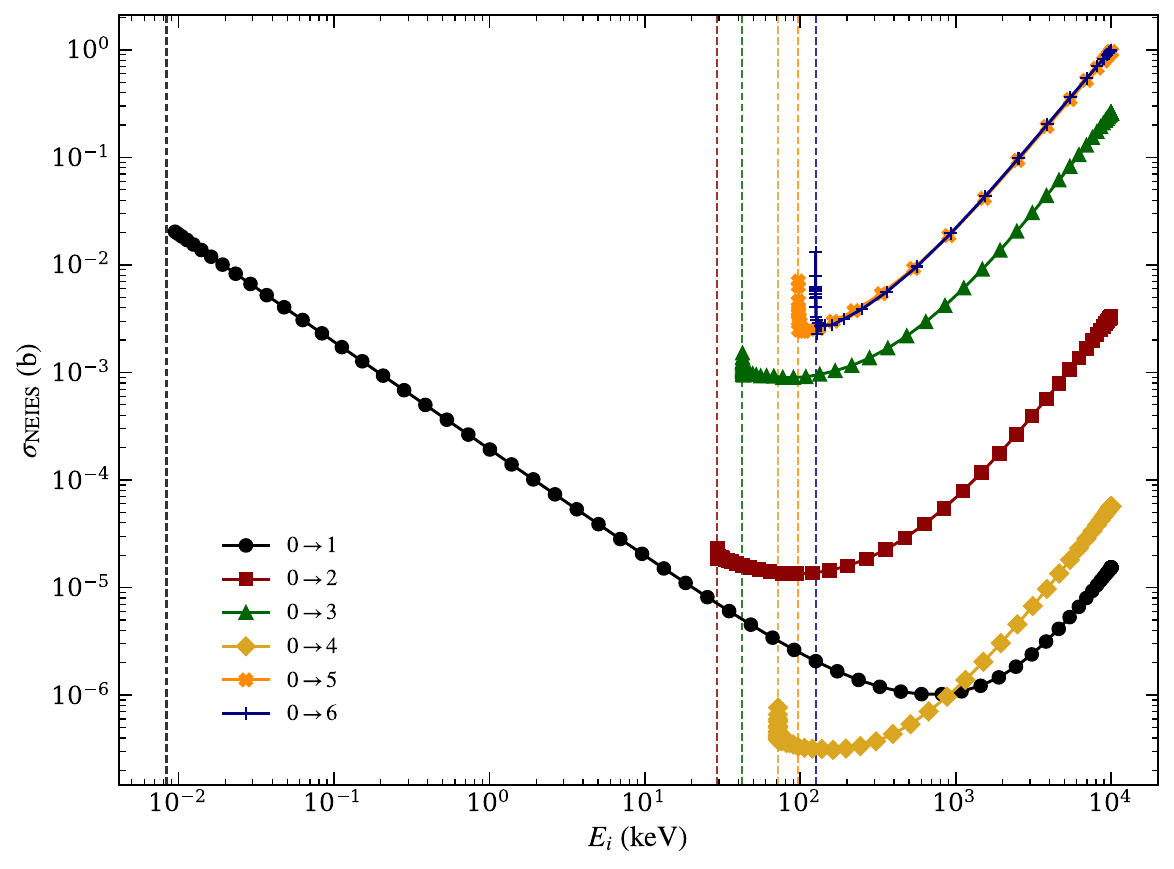}
\caption{\label{fig2_neies} 
NEIES cross sections $\sigma_{\mathrm{NEIES}}$ as functions of the incident electron energy $E_i$ for different reaction channels at an ionic charge state of $q_i = 90$. The vertical dashed lines mark the threshold energies corresponding to different excitation channels $0\rightarrow\beta$. 
}
\end{figure}

For a given incident electron energy, the NEIES cross section exhibits relatively small sensitivity to the ionic charge state, varying by approximately one order of magnitude at the same energy\cite{PhysRevC.106.044604, PhysRevC.110.064621}. 
In this work, we mainly focus on the NEIES cross section for the highest initial ionic charge state $q_i=90$, {i.e. $^{229}$Th$^{90+}$ }, which generally corresponds to the largest cross-section values among different excitation channels for a given ionic charge state and electron energy, as illustrated in Fig.~\ref{fig2_neies}. 
It is apparent that the cross sections exhibit significant energy-dependent characteristics: in the low-energy region (below around 200~keV), the cross section rapidly reaches its peak near the threshold energy $E_n$ and then decreases with increasing electron energy. Upon entering the high-energy region, the cross sections increase again and span several orders of magnitude; in some cases, the cross section for the fourth excited state even exceeds that of the first excited state, despite being smaller at lower energies. In addition, the cross sections for the fifth and sixth excited states exhibit a pronounced absolute advantage, with values far exceeding those for the direct excitation of the first excited state.

\begin{figure}[!htbp] 
\centering
\includegraphics[width=\linewidth]{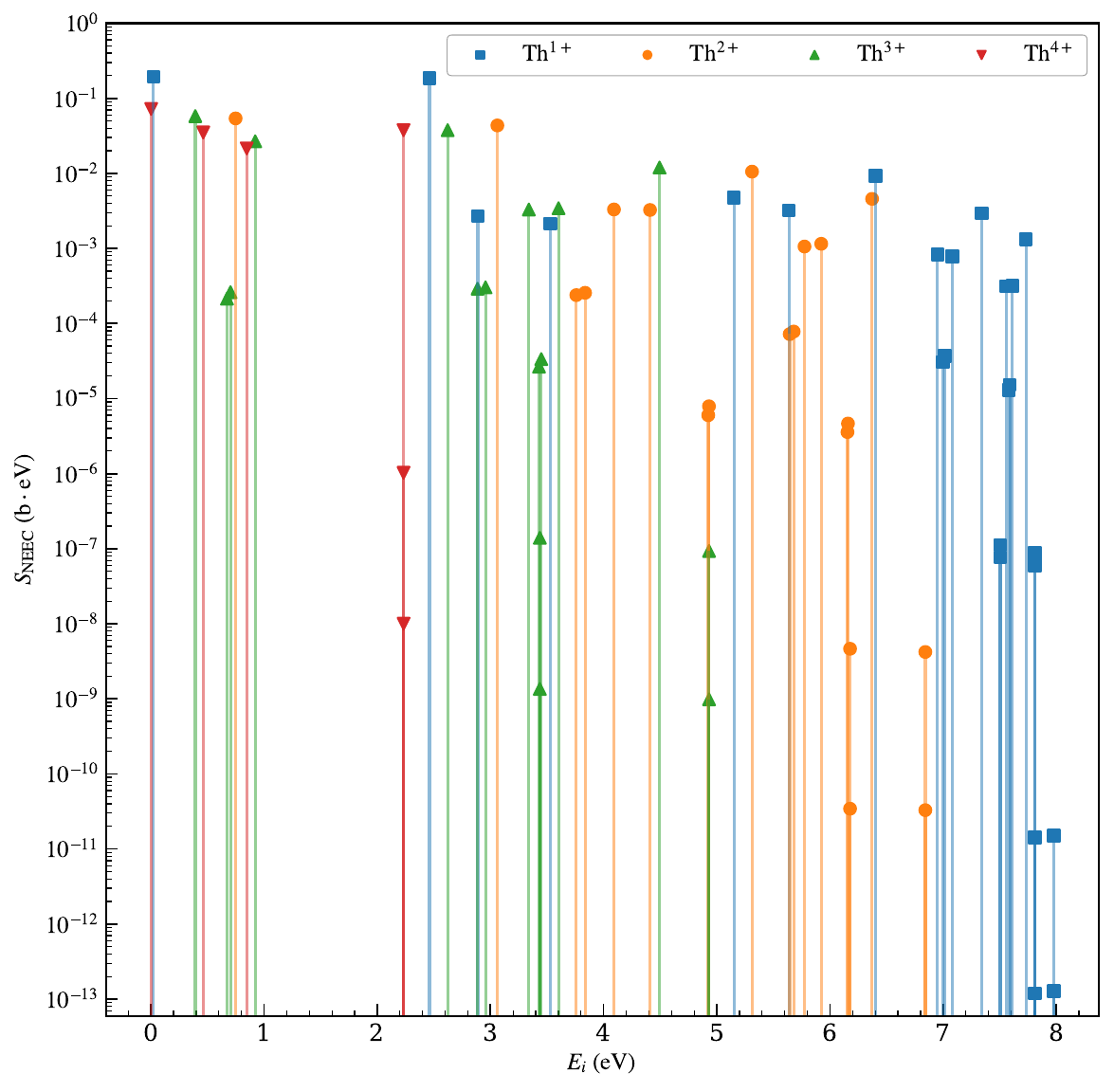}
\caption{\label{fig1_neec} 
NEEC resonance strength $S_{\mathrm{NEEC}}$ for the $0\!\rightarrow\!1$ excitation channel at $q_i=1$--4 when considering $n < 10$.
}
\end{figure}

Regarding the NEEC process, the cross-sections exhibit resonant structures.
In this work, the first excited isomer is discussed separately from higher states due to its much lower excitation energy, on the order of eV, compared with the keV range of the higher states.
This difference results in variations in the cross-section distributions.
For the first excited isomer at 8.36~eV, 
only the first few low-charge states satisfy the resonance condition when considering $n < 10$, as shown in Fig.~\ref{fig1_neec}.
If higher-$n$ Rydberg states, typically with $n \gtrsim 100$, are included~\cite{xu2025charge}, high-charge states can also contribute. 
Under ideal conditions, the maximum NEEC resonance strength $S_\mathrm{NEEC}$ reaches approximately 0.1~b$\cdot$eV near the resonance electron energy $E_{r} \simeq 0.1$~eV. 
In practice, however, such high-$n$ states may be difficult to realize experimentally with sufficient precision.

\begin{figure}[!htbp]
\centering
\includegraphics[width=\linewidth]{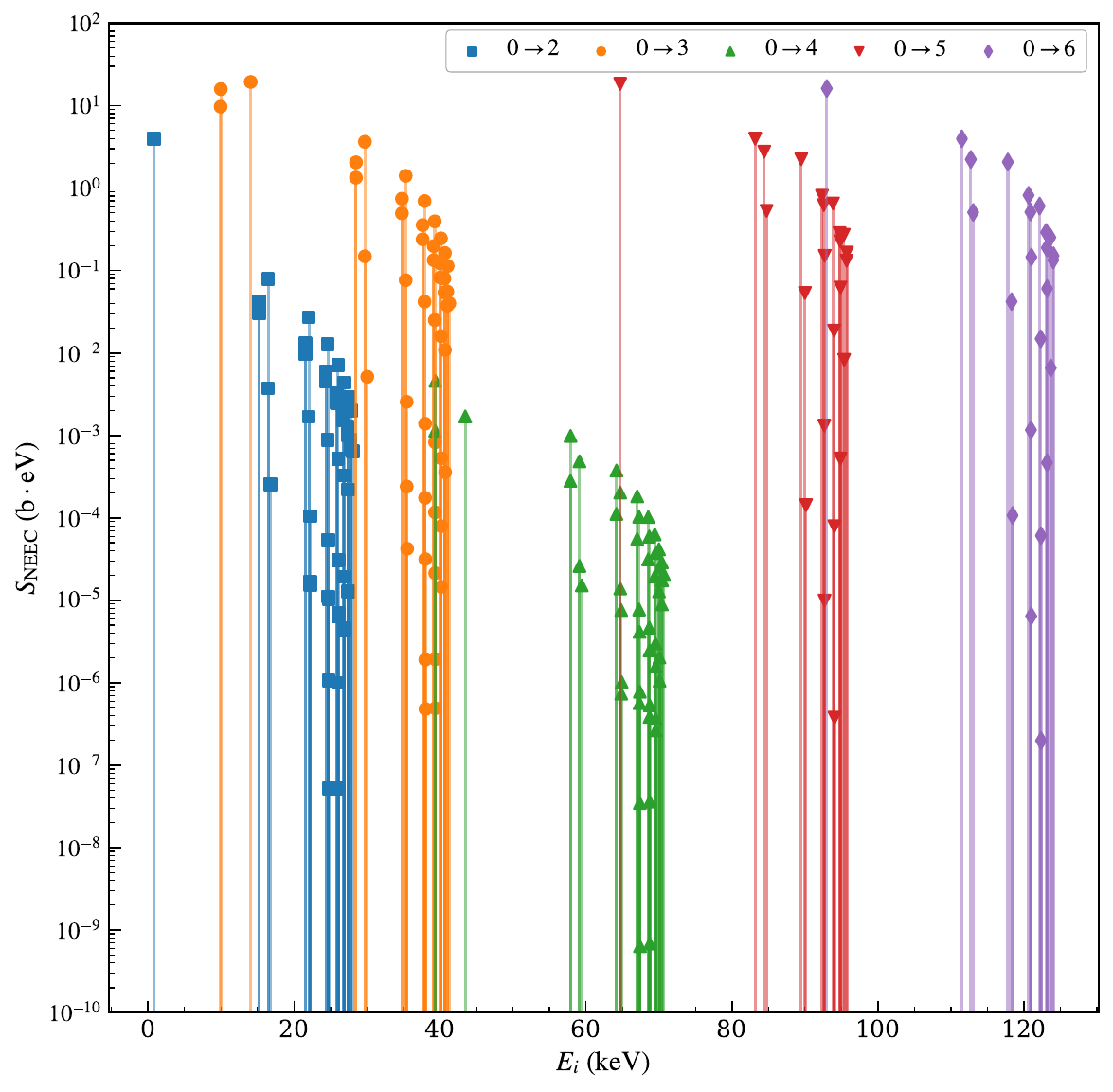}
\caption{ \label{fig3_neec} 
NEEC resonance strength $S_{\mathrm{NEEC}}$ for the $0\!\rightarrow\!2$--$6$ excitation channels at a selected representative ionic charge state $q_i=90$.
}
\end{figure}

For the second excited state and higher nuclear states, under the ground-state assumption~\cite{gargiulo2022nuclear}, the highest charge state satisfying the NEEC resonance condition generally corresponds to the global maximum of the cross section among all charge states~\cite{zhao2024efficient, xu2025charge}.
Even if this does not coincide with the absolute peak, it typically remains among the largest values. 
Accordingly, we focus on the case expected to yield the maximum NEEC resonance strength. 
The ionic charge state $q_i = 90$ is taken as an example for each excitation channel, with the results shown in Fig.~\ref{fig3_neec}. 
The largest $S_\mathrm{NEEC}$ values are concentrated in the third, fifth, and sixth excited states, reaching roughly 10~b$\cdot$eV. 
This is nearly two orders of magnitude higher than that of direct excitation.
Therefore, for NEEC excitation, indirect cascade excitation of \textsuperscript{229}Th appears to be a more suitable approach. 

\subsection{Yield rates in SRs}

In the SR environment, due to the presence of a single ionic charge state, for an electron beam at a given energy, the occurrence of NEIES precludes NEEC, and vice versa. Consequently, the total isomer production rate for all allowed cascade channels, defined as the number of \textsuperscript{229}Th isomers per unit time, can be expressed as
\begin{equation}
\begin{split}
R_{\mathrm{iso}} 
&= R_{\mathrm{NEEC}} +R_{\mathrm{NEIES}} \\
&= \left[ \lambda^{\mathrm{eff}}_{{\mathrm{NEEC}}}(E_{i}) + \lambda^{\mathrm{eff}}_{{\mathrm{NEIES}}}(E_{i})  \right] \times N_{i}.
\end{split}
\end{equation} 
Here, $N_{i} = n_i \pi r_i^2 L_i$ denotes the total number of target ions, where $n_i$ is the target ion density, and $r_i$ and $L_i$ define the radius and length of the ideal cylindrical interaction region, respectively.
For a typical SR~\cite{zhao2024efficient, PhysRevC.110.064621}, we assume a total of $10^8$ ions in the ring, and an electron beam with a current of $I_e = 200~\mathrm{mA}$, a radius of $r_e = 1~\mathrm{cm}$, and an energy spread on the order of 1~meV.

\begin{figure}[!htbp]
\centering
\includegraphics[width=\linewidth]{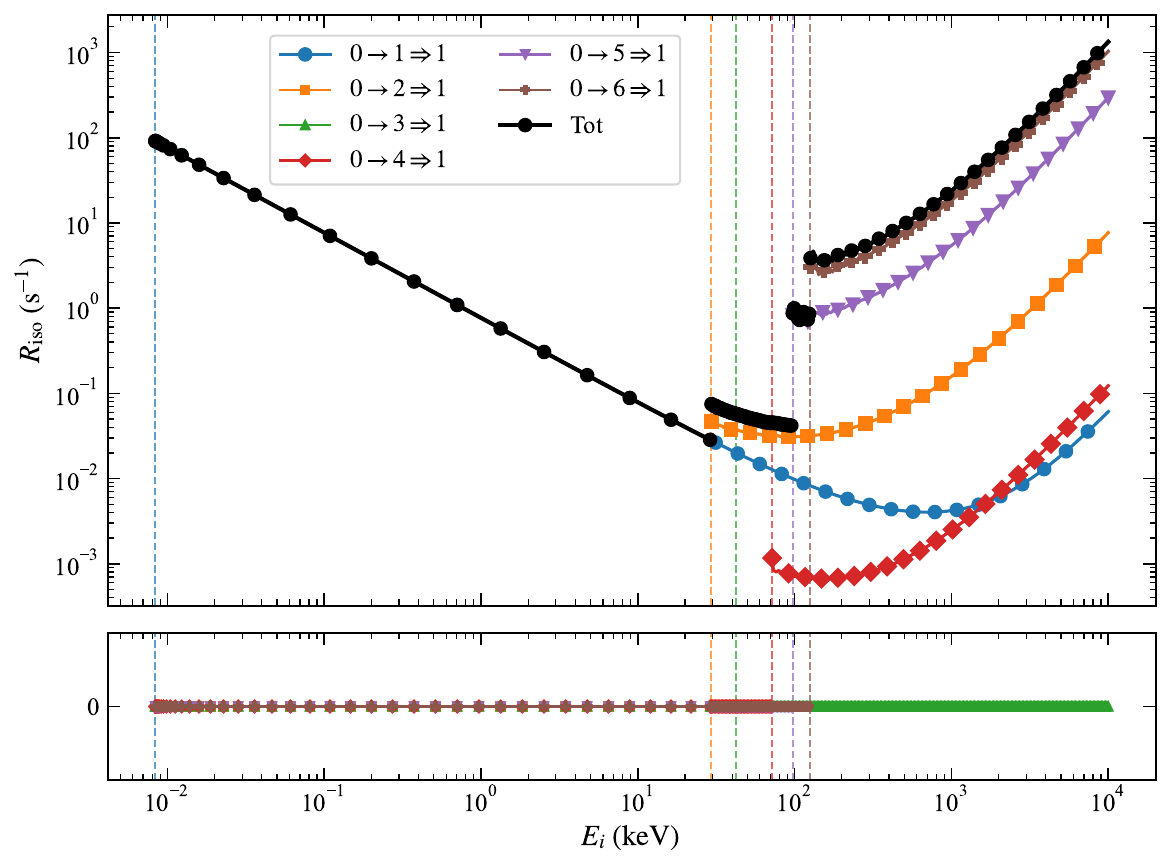}
\caption{\label{fig_neise_rate_SR_log} 
\textsuperscript{229m}Th NEIES production rates $R_{\mathrm{iso}}$ for different cascade decay channels as shown in Fig.~\ref{fig1-levels}  
and the total yield rate versus electron energy at $q_i = 90$. Vertical dashed lines denote excitation thresholds. A SR configuration with $10^{8}$ ions, $I_{e}=200~\mathrm{mA}$, $r_{e}=1~\mathrm{cm}$, and $\epsilon_e=1~\mathrm{meV}$ electron energy spread is assumed.}
\end{figure}

For the contribution due to the NEIES mechanism, the calculated isomer production rates are shown in Fig.~\ref{fig_neise_rate_SR_log}. As the electron energy increases, more excitation channels become accessible, leading to stepwise rises in the total yield rates at low energies. Within the energy range considered here, the total yield rates follow a trend similar to, but slightly higher than, that of the sixth excited state. Compared to the direct excitation of the first excited state, the total yield rates at high electron energies are enhanced by at least 4 orders of magnitude. Finally, since the branching ratio of the third excited state is zero, it does not contribute to the total yield rates across the entire energy range.

In the case of the NEEC process, the maximum yield rate for each cascade decay channel occurs at the electron beam energy corresponding to the highest NEEC cross section among all charge states and configurations. Table~\ref{R_isomer_neec_SR} lists, for each decay channel, the electron energies at which the NEEC cross section reaches its maximum for $q_i = 90$, along with the resulting isomer population rates for different levels $\beta$. Contributions from NEEC and NEIES are shown separately. The cross sections for direct isomer excitation are estimated following Ref.~\cite{xu2025charge}, although achieving these would require very high principal quantum numbers $n$. Notably, the absence of NEEC signals for the third excited state provides a convenient reference for control experiments. The cascade decay channel of the fourth excited state does not lead to an increased yield due to its relatively low cross section. The remaining cascade decay channels exhibit at least an order-of-magnitude enhancement relative to direct excitation, with the sixth excited state showing the largest increase, reaching approximately a factor of 80.

\begin{table*}[!htbp]  
\centering
\caption{\textsuperscript{229m}Th isomer population rates in a SR for different indirectly excited levels $\beta$.
$R_{\mathrm{iso}}$ denotes the total isomer yield rate, $R_{\mathrm{NEEC}}$ the contribution from NEEC, $R_{\mathrm{NEIES}}$ the contribution from NEIES, and $P$ the relative yield rate of each indirect excitation scheme compared with direct excitation.
}
\label{R_isomer_neec_SR}
\begin{ruledtabular}
\begin{tabular}{ccccccc}
Paths  & $E_{ir}$ (keV) & $R_{\mathrm{NEEC}} $ (s$^{-1}$) & $R_{\mathrm{NEIES}} $ (s$^{-1}$) & $R_{\mathrm{iso}}$ (s$^{-1}$) & Dominated by & $P$ \\
\hline
$0 \rightarrow 1$  & 0.0001     & 2.24$\times 10^{3}$  & 0.00                 & 2.24$\times 10^{3}$  & NEEC  & 1.00 \\
$0 \rightarrow 2 \rightarrow 1$  & 0.822871   & 5.16$\times 10^{4}$  & 9.34$\times 10^{-1}$ & 5.16$\times 10^{4}$  & NEEC  & 2.30$\times 10^{1}$ \\
$0 \rightarrow 3 \Rightarrow 1$ & 14.06784  & 0.00                  & 5.64$\times 10^{-2}$ & 5.64$\times 10^{-2}$ & NEIES & 2.52$\times 10^{-5}$ \\
$0 \rightarrow 4 \Rightarrow 1$ & 39.37434   & 5.61$\times 10^{1}$  & 5.99$\times 10^{-2}$ & 5.62$\times 10^{1}$  & NEEC  & 2.51$\times 10^{-2}$ \\
$0 \rightarrow 5 \Rightarrow 1$ & 64.68429   & 6.42$\times 10^{4}$  & 4.61$\times 10^{-2}$ & 6.42$\times 10^{4}$  & NEEC  & 2.87$\times 10^{1}$ \\
$0 \rightarrow 6 \Rightarrow 1$ & 92.98684   & 1.89$\times 10^{5}$  & 4.22$\times 10^{-2}$ & 1.89$\times 10^{5}$  & NEEC  & 8.44$\times 10^{1}$ \\
\end{tabular}
\end{ruledtabular}
\end{table*}

\subsection{Yield rates in EBITs}

In contrast to the SR case, the EBIT offers a plasma environment that supports the production and confinement of HCIs with a distributed charge-state population.
If only NEIES is considered, the weak dependence of the NEIES cross sections on the charge state leads to conclusions in an EBIT that are analogous to those in an SR.
For schemes incorporating NEEC resonances, a dual–electron-beam configuration~\cite{wang2023feasibility,zhao2024efficient} is adopted, in which one beam is used for charge breeding, while a second beam with tunable energy is dedicated to resonant NEEC excitation. Accordingly, the \textsuperscript{229m}Th production rate can be calculated by accounting for the charge-state distribution~$q_i$
\begin{equation}
\begin{split}
R_{\mathrm{iso}} 
&= R_{\mathrm{NEEC}} +R_{\mathrm{NEIES}} \\
&= \sum_{q_i, E_{ir}, E_{is}} P_{q_i} \left[ \lambda^{\mathrm{eff},~q_i}_{{\mathrm{NEEC}}}(E_{i}) + \lambda^{\mathrm{eff},~q_i}_{{\mathrm{NEIES}}}(E_{i})  \right] \times N_{i},
\end{split}
\end{equation}
where $P_{q_i}$ is the population of \textsuperscript{229}Th ions in charge state $q_i$, $E_{is}$ is the electron beam energy used for charge stripping in the dual-electron-gun EBIT device, and $E_{ir}$ is the resonant electron beam energy required for the NEEC process. Assuming an EBIT trapping volume approximated as a perfect cylinder with a radius equal to the electron beam radius, $50~\mu\mathrm{m}$, and a length of $3~\mathrm{cm}$, together with a target ion density of $10^8~\mathrm{cm^{-3}}$, the total number of target particles is estimated to be $N_{i} = 2.36 \times 10^{4}$.

\begin{figure}[!htbp]
\centering
\includegraphics[width=1\linewidth]{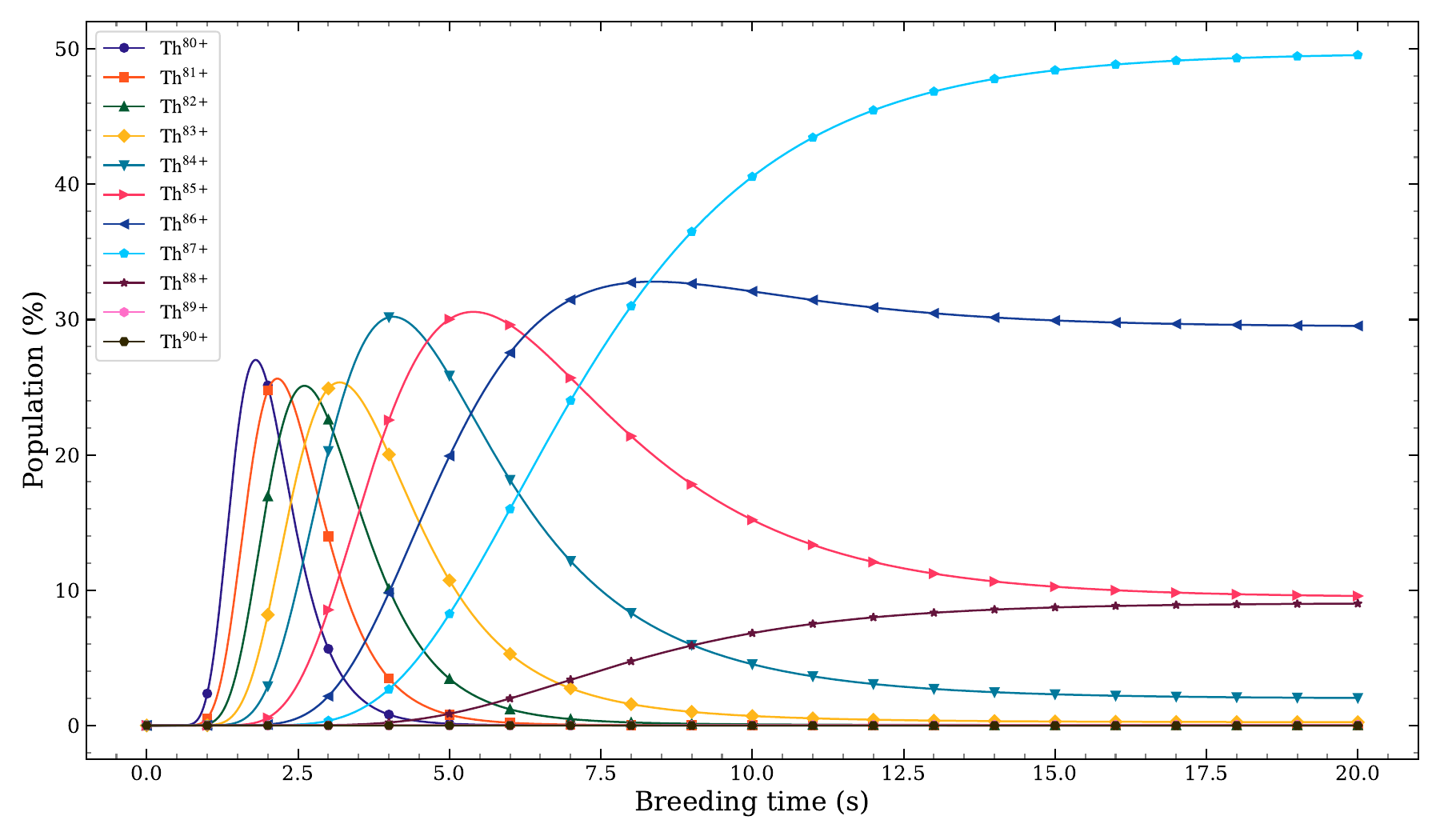}
\caption{\label{fig_ebitsim}
The charge-breeding process of  \textsuperscript{229}Th ions in an EBIT was simulated using the code \texttt{ebitsim}, with the following parameters: electron stripping energy of 130~keV, electron beam current of 200~mA, breeding time of 20~s, and beam radius of $50~\mu\mathrm{m}$.
}
\end{figure}

In this work, the time evolution simulation of ionic charge states within the trap is modeled using the charge-breeding simulation package \texttt{ebitsim}~\cite{ebitsim2025}. The code solves a system of coupled rate equations describing the population dynamics of different ionic charge states under electron-beam irradiation. Building on earlier developments, \texttt{ebitsim} allows flexible control of key EBIT parameters, including the electron-beam current and energy, beam diameter, background gas pressure, and the spatial overlap between the electron beam and the trapped ion cloud. 

For the charge-breeding simulations of \textsuperscript{229}Th ions, the dominant first-order atomic processes are taken into account, namely electron-impact ionization (EI), radiative recombination (RR), and charge exchange (CX) with residual gas atoms. EI drives the population toward higher charge states, whereas RR and CX provide competing loss channels that limit the achievable charge-state distribution. Higher-order processes, such as dielectronic recombination, ion-ion interactions, and nuclear-related processes (e.g., IC), are neglected. This approximation is justified under typical EBIT operating conditions, where the electron density and background gas pressure are sufficiently low. Within this framework, \texttt{ebitsim} provides a self-consistent description of the temporal evolution toward charge-state equilibrium and enables systematic exploration of how EBIT operating parameters influence the charge-breeding efficiency and the resulting charge-state distribution of \textsuperscript{229}Th ions.
For ions in a given charge state $q_i$, the time evolution of the charge-state population in the trap is governed by
\begin{equation}
\frac{d N_{q_i}}{d t}
= \mathrm{EI}_{q_i-1} - \mathrm{EI}_{q_i}
+ \mathrm{RR}_{q_i+1} - \mathrm{RR}_{q_i}
+ \mathrm{CX}_{q_i+1} - \mathrm{CX}_{q_i} ,
\end{equation}
which follows the rate-equation approach adopted in Ref.~\cite{wallis2021optimising}. In this code, ion loss from the trap is neglected due to the deep trapping potential characteristic of EBIT operation. 

As shown in Fig.~\ref{fig_ebitsim}, the case with a charge-stripping electron energy of 130~keV, a 200~mA electron beam, a breeding time of 20~s, a beam radius of $50~\mu\mathrm{m}$, and all other parameters set to their default values was simulated.
Subsequently, as the simulation indicates that the maximum charge-state population occurs at $q_i = 87+$, a second electron beam providing the resonant energies $E_{ir}$ corresponding to the maximum NEEC cross sections of each decay channel at $q_i = 87+$ is employed. The results are summarized in Table~\ref{R_isomer_130keV_EBIT}. 
It is found that the third and fourth excited states are more favorable for NEIES dominance, as the maximum NEEC cross sections along these cascade decay paths remain relatively small, while NEIES contributes across the full energy range. Other cascade paths, except for the sixth excited state, are in principle not suitable to distinguish NEEC from NEIES, since the theoretical difference between the two mechanisms is only about one order of magnitude. Moreover, for the sixth cascade path, under 130~keV stripping electron energy and 94.6615~keV resonance electron energy, it provides the largest advantage compared to direct excitation. Here, NEEC dominates and can produce \textsuperscript{229}Th isomeric states at a rate of 13.5~s$^{-1}$, resulting in a total \textsuperscript{229m}Th yield rate that is 29 times higher than that from direct excitation. In summary, this cascade decay channel enables not only efficient production of \textsuperscript{229m}Th but also offers a promising route for the experimental verification of the NEEC process.

\begin{table*}[!htbp] 
\centering
\caption{\textsuperscript{229m}Th isomer population rates for an EBIT with electron stripping energy $E_{is} = 130~\mathrm{keV}$.
$E_{ir}$ denotes the resonance electron energy input corresponding to the maximum NEEC cross section for each decay channel.  
Separate contributions from NEEC and NEIES are listed.
$R_{\mathrm{iso}}$ denotes the total isomer yield rate, $R_{\mathrm{NEEC}}$ the contribution from NEEC, $R_{\mathrm{NEIES}}$ the contribution from NEIES, and $P$ the relative yield rate of each indirect excitation scheme compared with direct excitation.}
\label{R_isomer_130keV_EBIT}
\begin{ruledtabular}
\begin{tabular}{cccccccc}
Paths &
$E_{is}$ (keV) &
$E_{ir}$ (keV) &
$R_{\mathrm{NEEC}}$ (s$^{-1}$) &
$R_{\mathrm{NEIES}}$ (s$^{-1}$) &
Dominated by &
$R_{\mathrm{iso}}$ (s$^{-1}$) &
$P$ \\
\hline
$0 \rightarrow 1$ & 130 & 0.0001    & 2.09$\times10^{-1}$ & 2.64$\times10^{-1}$ & NEEC and NEIES & 4.73$\times10^{-1}$ & 1.00 \\
$0 \rightarrow 2 \rightarrow 1$ & 130 & 2.519733  & 1.31                & 2.84$\times10^{-1}$ & NEEC and NEIES & 1.59                & 3.37 \\
$0 \rightarrow 3 \Rightarrow 1$ & 130 & 15.764703 & 0.00                & 2.67$\times10^{-1}$ & NEIES          & 2.67$\times10^{-1}$ & 0.56 \\
$0 \rightarrow 4 \Rightarrow 1$ & 130 & 41.049    & 4.31$\times10^{-3}$ & 2.69$\times10^{-1}$ & NEIES           & 2.73$\times10^{-1}$ & 0.58 \\
$0 \rightarrow 5 \Rightarrow 1$ & 130 & 66.35895  & 4.20                & 2.68$\times10^{-1}$ & NEEC and NEIES & 4.47                & 9.45 \\
$0 \rightarrow 6 \Rightarrow 1$ & 130 & 94.6615   & 1.35$\times10^{1}$  & 2.67$\times10^{-1}$ & NEEC & 1.38$\times10^{1}$  & 29.18 \\
\end{tabular}
\end{ruledtabular}
\end{table*}

\section{Conclusions}\label{sec_summary}

In conclusion, to excite  \textsuperscript{229m}Th efficiently, we propose electron-induced excitation schemes based on two typical experimental platforms, SRs and EBITs.
These schemes leverage the large excitation cross sections of HCIs and the high efficiency of subsequent cascade decays, which are expected to significantly enhance the yield rate of the \textsuperscript{229m}Th isomer.
The numerical results show that, under the NEIES mechanism across a continuous range of electron energies, the \textsuperscript{229m}Th yield rate at electron energies above approximately 200~keV can be significantly enhanced. This enhancement occurs along the cascade decay pathways of the fifth and sixth excited states and can exceed that from direct excitation by up to four orders of magnitude.        
Furthermore, at the NEEC resonance energy, the yield rate can be further increased by dozens of times.      

Once a substantial number of \textsuperscript{229m}Th ions are produced, they can be extracted and loaded into various ion traps, providing valuable input for the development of nuclear clocks. 
Alternatively, they can be embedded in crystal environments to study their decay modes. 
The proposed schemes also highlight the potential of using \textsuperscript{229}Th in SR and EBIT facilities to investigate the NEEC process.

\begin{acknowledgements}
This work is supported by the National Key Research and Development Program of China (Grant No.~2023YFA1606900), the National Natural Science Foundation of China (Grant Nos.~12235003 and 12447106), and the China Scholarship Council (Grant No.~202406100161).

\end{acknowledgements}

\nocite{*}

\hypersetup{urlcolor=blue}
\bibliography{refs}  

\end{document}